\documentclass{article}

\PassOptionsToPackage{numbers, compress}{natbib}

\usepackage[preprint]{neurips_2023_ml4ps}

\usepackage[utf8]{inputenc} 
\usepackage[T1]{fontenc}    
\usepackage{hyperref}       
\usepackage{url}            
\usepackage{booktabs}       
\usepackage{amsfonts}       
\usepackage{nicefrac}       
\usepackage{microtype}      
\usepackage{xcolor}         
\usepackage{amsmath} 
\usepackage{graphicx}
\usepackage{subfigure}

\title{GAMMA: Galactic Attributes of Mass, Metallicity, and Age Dataset}

\author{%
  Ufuk Çakır\\
  Interdisciplinary Center\\ for Scientific Computing,\\
  University of Heidelberg,\\
  Im Neuenheimer Feld 205,\\ D-69120 Heidelberg\\
  \texttt{ufuk.cakir@stud.uni-heidelberg.de} \\
    \And
     Tobias Buck\\ 
  Interdisciplinary Center\\ for Scientific Computing,\\  University of Heidelberg,\\ Im Neuenheimer Feld 205,\\ D-69120 Heidelberg\\
  \texttt{tobias.buck@iwr.uni-heidelberg.de} \\
}

\begin{document}

\maketitle

\begin{abstract}
We introduce the \textbf{GAMMA} (\textbf{G}alactic \textbf{A}ttributes of \textbf{M}ass, \textbf{M}etallicity, and \textbf{A}ge) dataset, a comprehensive collection of galaxy data tailored for Machine Learning applications. 
This dataset offers detailed 2D maps and 3D cubes of 11 727 galaxies, capturing essential attributes: stellar age, metallicity, and mass. 
Together with the dataset, we publish our code to extract any other stellar or gaseous property from the raw simulation suite to extend the dataset beyond these initial properties, ensuring versatility for various computational tasks. Ideal for feature extraction, clustering, and regression tasks, \textbf{GAMMA} offers a unique lens to explore galactic structures using computational methods and is a bridge between astrophysical simulations and the field of scientific machine learning (ML).
As a first benchmark, we applied Principal Component Analysis (PCA) to this dataset.
We find that PCA effectively captures the key morphological features of galaxies 
with a small number of components. We achieve a dimensionality reduction by a factor of $\sim$200 ($\sim$3 650) for 2D images (3D cubes) with a reconstruction accuracy below 5\%. 

\end{abstract}

\section{Motivation}

Galactic structures have long been one of the cornerstones of astrophysics. As computational methods, particularly ML, gain prominence, the need for datasets that bridge astrophysical simulations and computational techniques is growing. 
The \textbf{GAMMA} dataset provides a well-curated collection of 2D and 3D galaxy images, which captures crucial physical attributes for understanding galaxy morphology: stellar age, metallicity, and mass. It combines well-understood physical relations in the data with explicit and controlled image generation processes such as noise level or point-spread function. As such, this dataset not only offers a unique opportunity for feature extraction, clustering, and regression tasks, but also paves the way for a deeper understanding of galactic structures through computational lenses. The dataset is available on Zenodo.\footnote{\url{https://zenodo.org/records/8375344}} Additionally, we publish all of our code on GitHub\footnote{\url{https://github.com/ufuk-cakir/GAMMA}}, which allows one to easily extract images from raw simulations or create higher resolution realizations of the data.  
In this way, we facilitate the community to go beyond the exploratory benchmarks presented in this work.


\section{Related Work}
\label{gen_inst}

Hydrodynamical simulations of galaxy formation such as IllustrisTNG \citep{Pillepich_2017}, EAGLE \citep{Schaye2015} CAMELS \citep{camels} or zoom-in models such as AURIGA \citep{Grand2017}, FIRE \citep{Hopkins2018}, VINTERGATAN \citep{Agertz2021} or NIHAO \citep{Wang2015,Buck2020} generate extensive, multidimensional datasets that include various physical properties of galaxies. Similarly, the GALAXY ZOO initiative \citep{galaxyzoo} has collected a large set of real multicolor galaxy images that have been classified in a citizen science project. These datasets serve as valuable resources for rigorously and quantitatively studying galaxy morphology. Combined with modern ML techniques, they enable the development of interpretable generative models for galaxy morphology \citep[e.g.][]{Lanusse2021}.    

For our initial benchmark with the \textbf{GAMMA} dataset, we employ PCA, which has been widely used in various fields for interpretable dimensionality reduction and feature extraction \citep{jolliffe_principal_2016_pca_review}, including the decomposition of face images into "eigenfaces" \cite{Turk_Pentland_eigenfaces_1991} and its adaptation to galaxy classification, where the basis vectors of the reduced space are termed "eigengalaxies" \cite{calleja_fuentes_first_eigengalaxies}. In the context of galaxy morphology, \cite{Uzeirbegovic_2020} recently used PCA to transform a set of galaxy images into a space where "closeness" corresponds to visual similarity. 
Expanding upon these prior studies, we explore PCA's utility in jointly modeling the distribution of mass, metallicity, and stellar age in both two and three dimensions. This approach condenses the high-dimensional distribution of galaxies into a lower-dimensional space that retains essential morphological features and facilitates interpretable analysis. 

\section{GAMMA: A dataset for stellar mass, metallicity, and age maps}
\label{gamma}

\begin{figure}
   \subfigure[2D]{
       \includegraphics[width=.49\hsize, trim={0 16cm 0 1.5cm},clip]{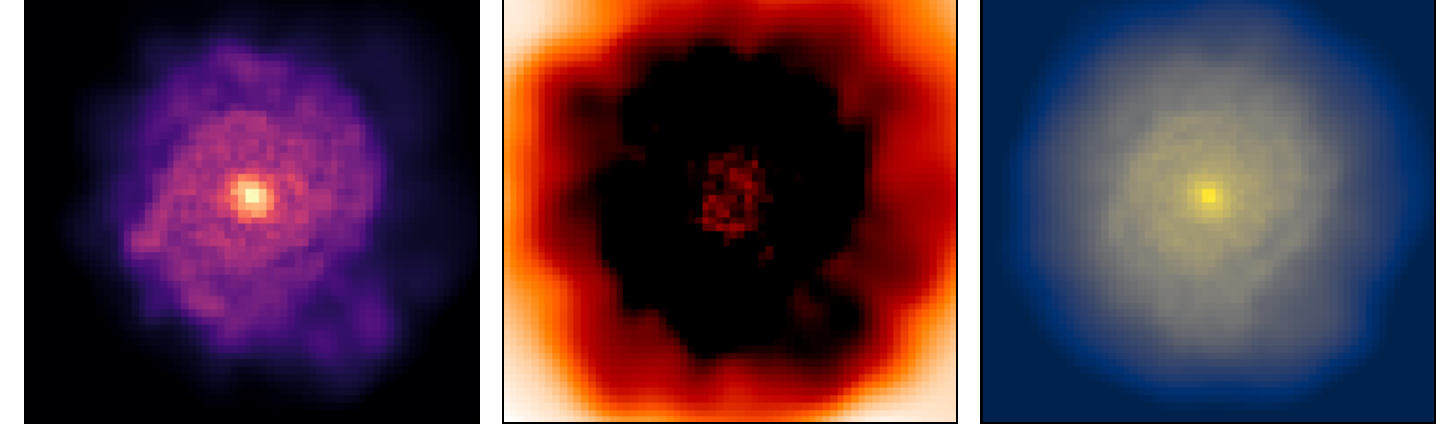}
  }
   \subfigure[3D]{
      \includegraphics[width=.49\hsize, trim={0 16cm 0 1.5cm},clip]{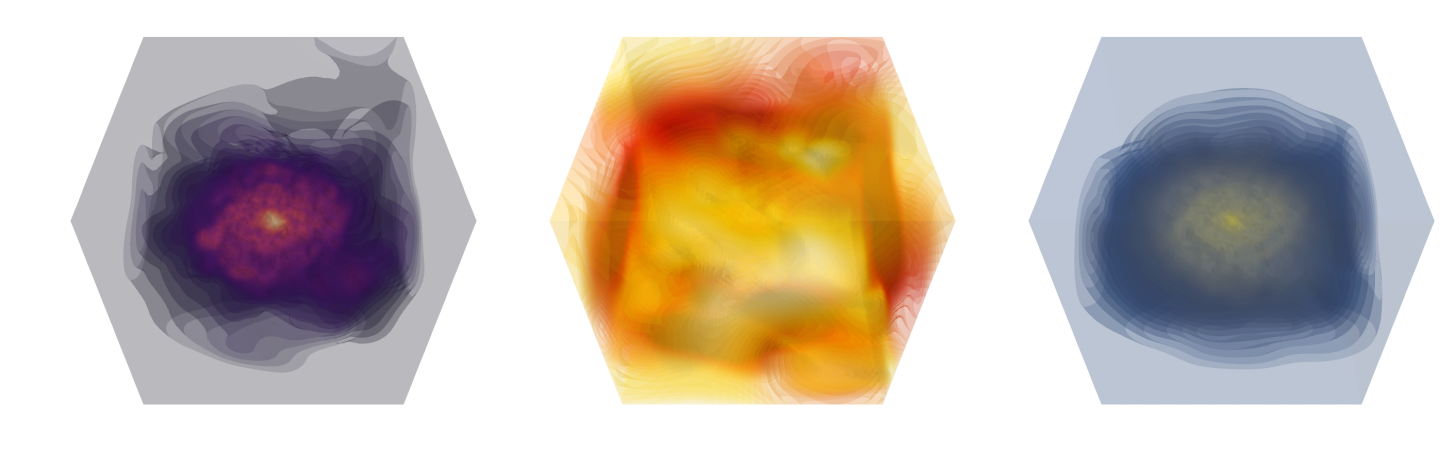}
   }
   \vspace{-.5cm}
   \caption{Each row showcases a sample galaxy represented in three stellar maps: metallicity, age, and mass from left to right. The pixel values have been normalized to the range of $[0,1]$, where brighter pixels correspond to higher values. Note that in the stellar age maps, dark pixels represent young stars. The three-dimensional plots were generated using the Python package \emph{plotly} \citep{plotly}.}
   \label{fig: sample_galaxies_after_prepocessing}
\end{figure}

We extracted galaxy data from the publicly available IllustrisTNG-100 simulation suite\footnote{\url{https://www.tng-project.org}} \citep{Nelson_2017,Pillepich_2017,Springel_2017} specifically from snapshot $99$ (redshift $z=0$) within the mass range $10^{9.5}M_\odot/h < M_\star < 10^{13}M_\odot/h$. We excluded galaxies with \texttt{SubhaloFlag}$=0$ and those without particles on the edges, producing a total of $N=11727$ galaxies. Particles with negative \texttt{GFM\_StellarFormationTime} are masked out, as they correspond to wind-phase gas cells and hence are not relevant for our analysis.
For uniform rotation, each galaxy undergoes face-on rotation based on the eigenvectors of the moment of inertia tensor and a further tilt correction using PCA with two components to align the semi-major axis of the galaxy with the x-axis.
Using the \textit{SWIFTsimIO} library's render module \citep{Borrow2020} the 2D and 3D images were generated, properly reflecting the physical and spatial resolution by employing the smoothing length parameter.
These images have $64^2$ and $64^3$ pixel resolutions and depict \texttt{particle mass}, \texttt{mass-weighted metallicity} and \texttt{stellar age}, spanning five stellar half-mass radii in all directions.
Pixel values are normalized to $[0, 1]$ and adjusted with $25^\mathrm{th}$ percentile clipping to standardize intensities and reduce noise-induced variance.
The final images for all fields are stored in HDF5 format, whereas the \texttt{SubhaloId} and stellar mass values for each galaxy are located in the \textit{attributes} subgroup. In Fig.~\ref{fig: sample_galaxies_after_prepocessing} we show a typical face-on spiral galaxy in the dataset.

\section{Benchmark: Principal Component Analysis on Images and Data Cubes}
\label{pca}

\begin{figure}
    \includegraphics[width=.33\hsize]{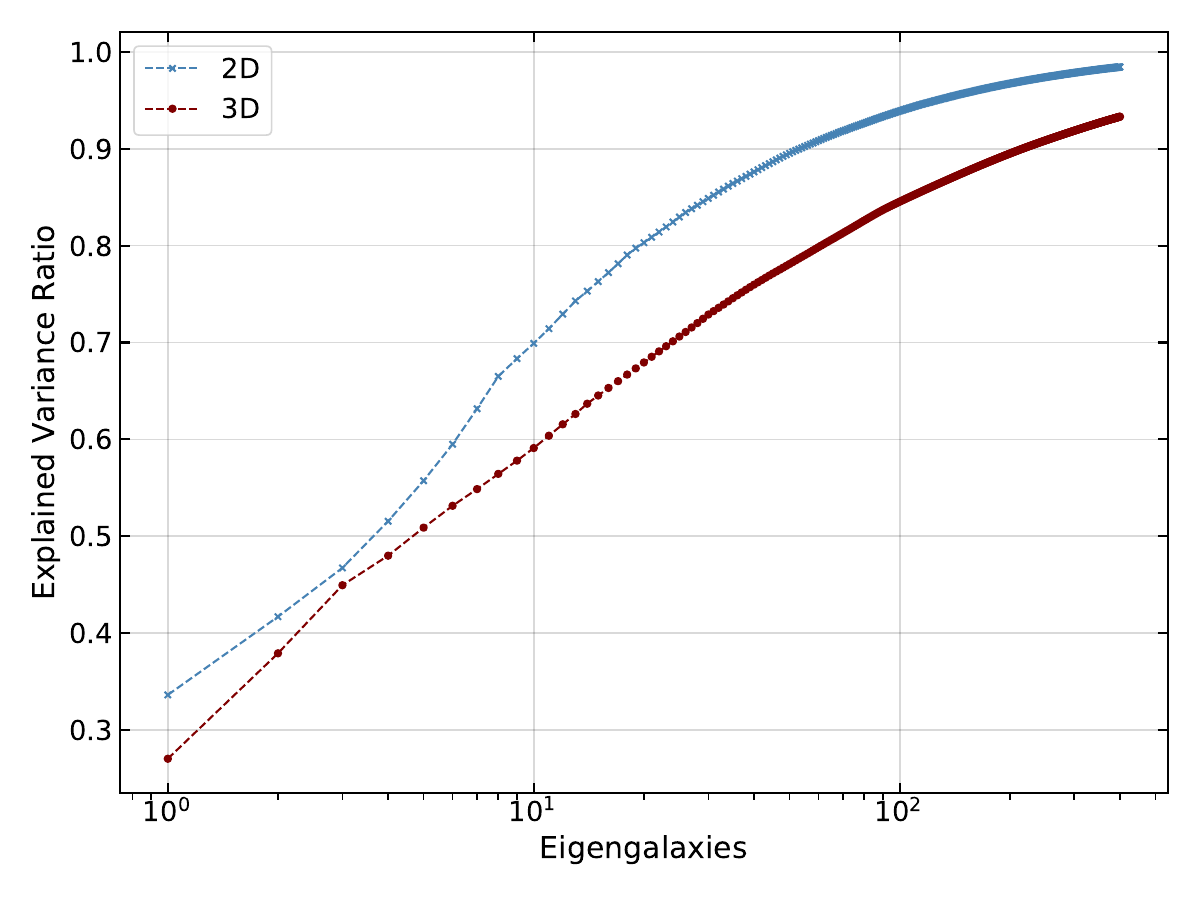}
    \includegraphics[width=.33\hsize,trim={0.1cm .1cm 0 0},clip]{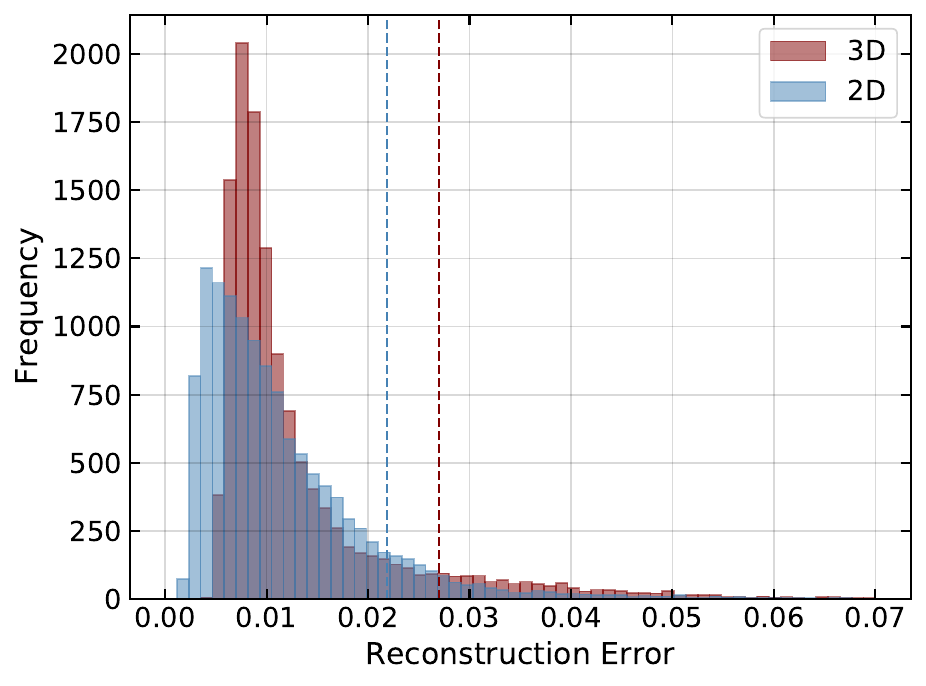}
    \includegraphics[width=.33\hsize,trim={0.1 .1cm 0 0.cm},clip]{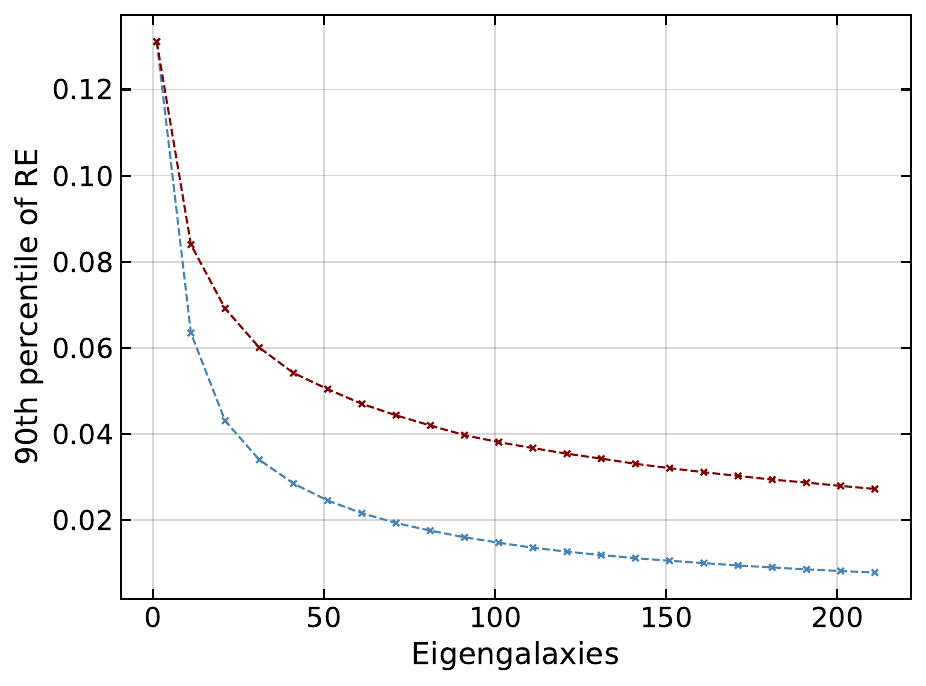}
    \vspace{-.6cm}
    \caption{\emph{Left:} Cumulative explained variance ratio (EVR) for up to 400 eigengalaxies: Achieving the same explained variance requires significantly more eigengalaxies in 3D compared to 2D. To surpass an 90\% EVR, $\sim60$ (215) eigengalaxies are needed in 2D (3D).
    \emph{Middle:} Reconstruction error (RE) for fixed-dimensionality reduction on 60 (215) eigengalaxies in 2D (3D). The dashed line represents the 90\% quantile. Impressively, 90\% of all images exhibit a RE below $0.022$ ($0.027$).
    \emph{Right:} 90th percentile of RE as a function of the number of eigengalaxies. Reconstruction is a strong function of eigengalaxies, and already 15 (60) eigengalaxies lead to a RE better than 5\% in 2D (3D).}   
    \label{fig:cumsum_explained_variance_ratio}
\end{figure}

\begin{figure*}
   \centering
   \subfigure[Spiral]{
      \includegraphics[width=.45\hsize]{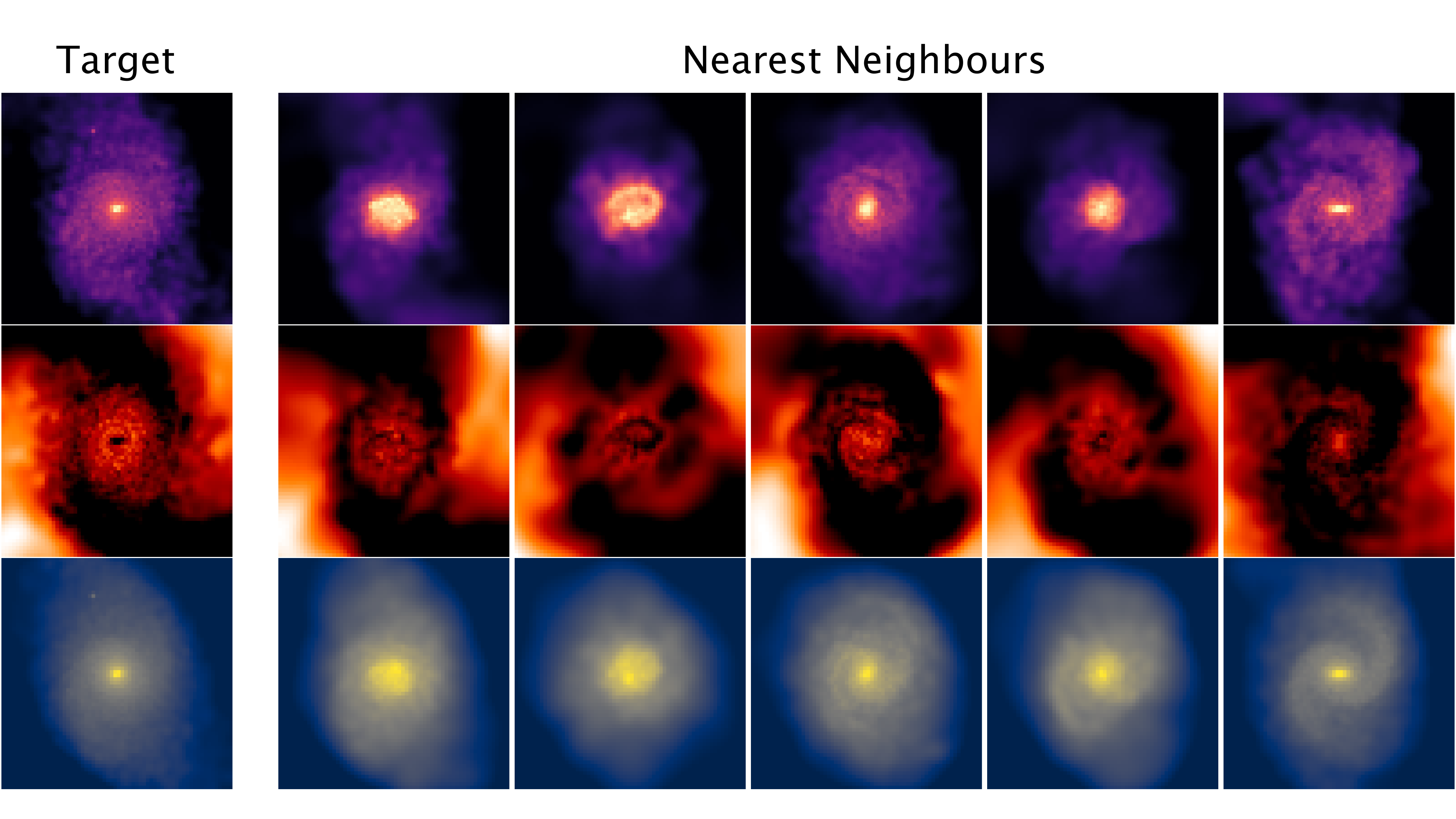} 
   }
   \subfigure[Bar]{
      \includegraphics[width=.45\hsize]{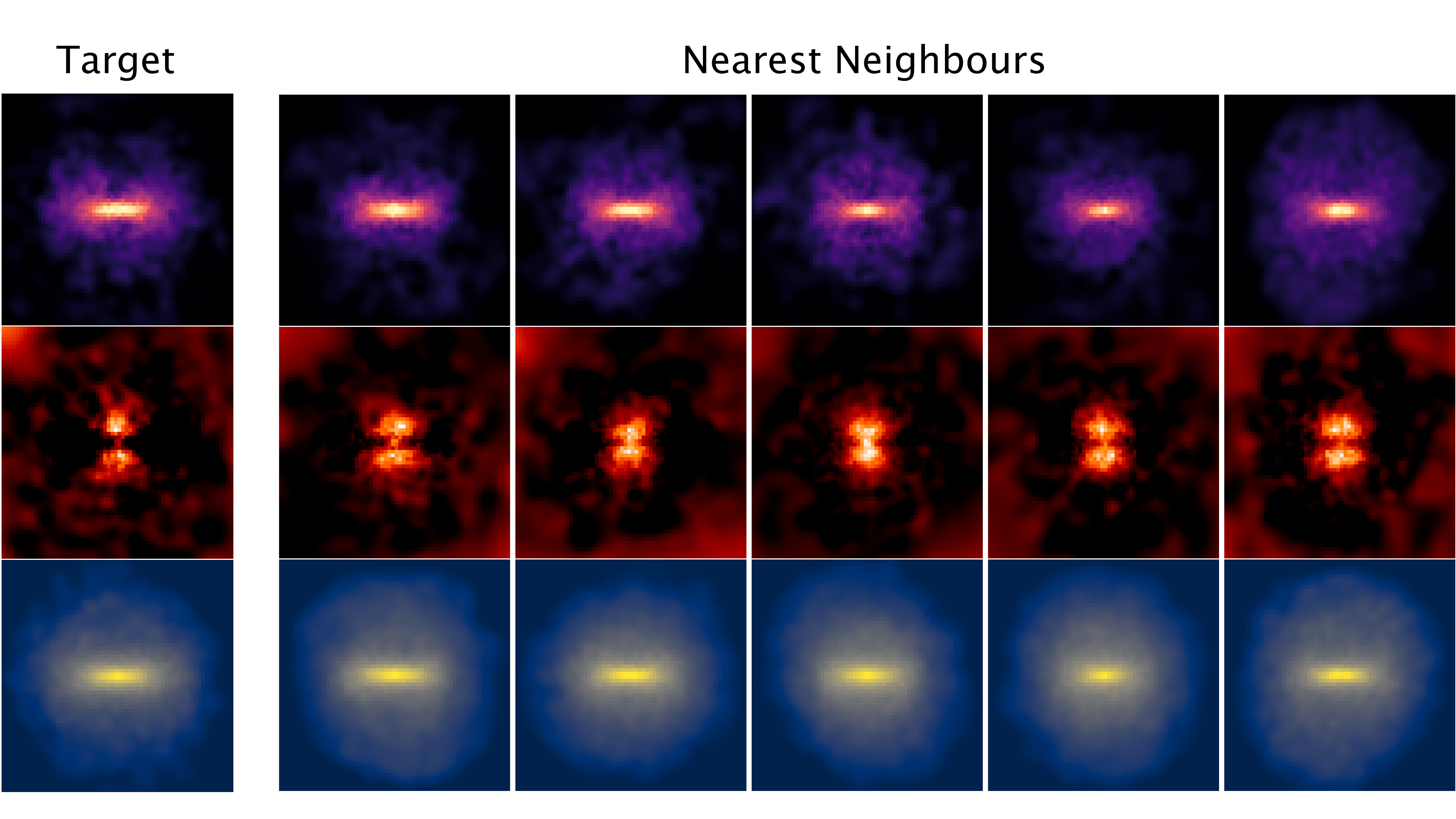}
   }
   \vspace{-.25cm}
   \caption{Nearest Neighbors in eigenspace: given a sample galaxy we find the five nearest neighbors in euclidean distance in the lower dimensional space (right) for a spiral (a) and bar shaped (b) galaxy.}
   \label{fig: nearest_neighbours}
\end{figure*}

We perform PCA in $\mathbb{R}^{N\times d}$ with $N=11727$ and $d=3\cdot64^2$ in 2D ($d=3\cdot64^3$ in 3D) using the \emph{scikit-learn} \citep{scikit-learn} Python package. PCA projects our high-dimensional data onto a lower-dimensional subspace spanned by the top $n$ eigengalaxies $\mathbf{\lambda_n} \in \mathbb{R}^{N\times d}$, preserving most of the variance. 
%
%
We evaluate the effectiveness of the PCA model in preserving information through the Explained Variance Ratio (EVR).
The cumulative sum of the EVR for up to 400 components is shown in Fig.~\ref{fig:cumsum_explained_variance_ratio}. 60 (215) eigengalaxies account for $\sim90\%$ of the total variance in 2D (3D), demonstrating substantial information retention despite a significant reduction in dimensionality by a factor of $205$ ($3641$). With as few as 16 eigengalaxies, major features such as spiral arms are well reconstructed. Increasing this number results in more accurate and finer image details. \\ To assess the robustness of the model, we perform a 400-fold cross-validation with 60 components on a 75\% random subsample of the data. The resulting EVR distribution has a mean of $0.9087$ and a standard deviation of only $0.0003$, demonstrating excellent robustness. This indicates that our dataset size is sufficient to encompass various morphologies and using 60 eigengalaxies robustly explains 90\% of the variance between morphologies in 2D.\\
We define the reconstruction error (RE) as the fractional difference in pixel values between the PCA representation, $\hat{\mathbf{I}}$, and the original image, $\mathbf{I}$ as 
\begin{displaymath}
\text{RE} = \frac{\sum_{k=1}^{d} (I_k-\hat{I}_k)^2}{\sum_{k=1}^{d} I_k}
\end{displaymath}
where $\hat{I}_k$ and $I_k$ represent the $k$-th elements (pixel) of the reconstructed and original vectors (images), respectively.  Using 60 (215) eigengalaxies, we calculate the RE and show its distribution in the middle panel of Fig.~\ref{fig:cumsum_explained_variance_ratio} for the 2D (blue) and 3D (red) cases. 
Remarkably, 90\% of all images have a RE below $0.022$ in 2D ($0.027$ in 3D), highlighting the accuracy of our PCA model. The right panel of Fig.~\ref{fig:cumsum_explained_variance_ratio} shows how the RE scales with the number of eigengalaxies. In the 2D case, as few as $\sim15$ eigengalaxies produce a RE below $0.05$ for 90\% of all images. In contrast, the 3D case requires $\sim60$ eigengalaxies to achieve a similar RE. These results are consistent with the EVR findings and demonstrate that a small number of eigengalaxies accurately describe galaxy morphology.\\
One of the significant advantages of PCA over more complex ML methods such as neural networks is its inherent interpretability. PCA provides a linear decomposition of the original image space, where each individual component can be interpreted as an image representing a specific morphology. We have visualized how eigengalaxies contribute to different morphologies. Notably, vastly different eigengalaxies contribute to the reconstruction of spiral and triaxial galaxies, as expected due to their distinct morphologies. For spiral galaxies, eigengalaxies with pronounced power along diagonal edges are essential for reconstructing the spiral arm feature. Conversely, for triaxial galaxies, eigengalaxies with nearly point-symmetry or bar-like features play a more prominent role in the reconstruction. This qualitative analysis underscores the potential of PCA decomposition in describing the galaxy morphology. Fitting galaxies with our PCA model enables the study of general morphological trends across different galaxies through the PCA scores of contributing eigengalaxies. For instance, it becomes straightforward to quantify and assess whether a galaxy exhibits bar-like, spiral-like, or centrally concentrated mass, age, or metallicity distributions. This interpretability is valuable for galaxy classification and similarity searches.\\
%
%
To assess whether the lower-dimensional image space encodes meaningful morphological information, we perform a morphological similarity search. This involves selecting a sample galaxy and searching for its nearest neighbors in the PCA eigenspace using the Euclidean distance metric: $d(\mathbf{\hat s}, \mathbf{s_i})=\|\mathbf{\hat s}-\mathbf{s_i} \|_2$.
Here, $\mathbf{\hat s}$ and $\mathbf{s_i}$ represent the scores of the sample and the remaining galaxies, respectively. In Fig.~\ref{fig: nearest_neighbours}, we show the five nearest neighbors for two different sample galaxies, which demonstrates the relationship between Euclidean proximity in the PCA eigenspace and morphologically similar characteristics. For the spiral galaxy (left panels), the five nearest neighbors exhibit pronounced two-arm spiral structures. For the triaxial galaxy (right panels), all nearest neighbors are also triaxial and exhibit a prominent bar-like feature in mass and metallicity maps. Additionally, PCA scores can be used for various other analyses, including clustering data using methods such as Gaussian mixture models \citep{Reynolds2009} or k-means \citep{Schoelkopf1998}, allowing unsupervised galaxy type classification. They can also be used for outlier detection, potentially identifying galaxies with unique and interesting morphological characteristics.

\section{Conclusions and Limitations}

We present the \textbf{GAMMA} dataset, a well-curated collection of 2D and 3D galaxy images that capture stellar age, metallicity, and mass. This dataset uniquely combines well-understood physical relations in the data with explicit and controlled image generation processes such as noise level or point-spread function. As such, this dataset paves the way for a deeper understanding of galactic structures and the public release of \textbf{GAMMA} will facilitate the community to go beyond the exploratory benchmarks presented in this work. So far, we have harnessed PCA to analyze galaxy images in both 2D and 3D, simultaneously modeling the distributions of mass, metallicity, and age. Our exploration has yielded several significant findings and insights into the potential applications of PCA for characterizing galaxy morphology and conducting morphological analysis:
\begin{enumerate}
      \item \textbf{Effective Reconstruction:} We have demonstrated that PCA can effectively reconstruct galaxy images with a relatively small number of eigengalaxies. The accuracy of these reconstructions was quantified using the reconstruction error metric, which measures the fractional difference in pixel values between the original and PCA-reconstructed images. Our results indicate that even with a modest number of 60 (215) eigengalaxies in 2D (3D), the PCA model achieves highly accurate reconstructions. Approximately 90\% of the images have reconstruction errors below $0.022$ ($\mathbf{0.027}$), underscoring the potential of PCA to efficiently represent the intricate morphology of the galaxy.
      
      \item \textbf{Interpretability:} PCA offers some unique advantage over more complex ML methods. PCA allows for a straightforward interpretation of its components, with each eigengalaxy representing a specific morphological feature. Visualizing the top contributing eigengalaxies for different types of galaxies reveals their distinctive contributions and demonstrates the potential of PCA for classifying galaxies based on their morphology.
      
      \item \textbf{Applications in Morphological Analysis:} We present the utility of PCA to conduct morphological similarity searches. By calculating Euclidean distances in the PCA eigenspace, we successfully identified galaxies that were morphologically similar to a given sample galaxy. This suggests that PCA captures meaningful features of galaxy morphology, enabling efficient similarity analysis and clustering. 

      \item \textbf{Limitations:} PCA is a linear decomposition. Each galaxy is reconstructed as a linear combination of eigengalaxies, so the initial unitary rotation of the dataset is important for PCA to find meaningful features. PCA assumes that the data can be understood in terms of orthogonal axes of maximum variance, which is why certain morphological features, such as rotational positions of spiral arms, may not be accurately fitted.
\end{enumerate}
In conclusion, our study underscores the potential of PCA as a powerful yet simple tool for analyzing and characterizing galaxy morphology. Its efficiency in image representation, interpretability of components, and facilitation of morphological similarity searches make it a valuable approach in the field of astrophysics. Future work could involve further refinement of the PCA model, such as exploring nonlinear PCA, investigating its applications in other astronomical datasets, and developing hybrid approaches that combine PCA with other ML techniques such as variational autoencoders, GANs, or diffusion models for even more comprehensive analyses of galaxy images. It could also be valuable to explore benefits from including simulation data from other cosmological simulations such as EAGLE, CAMELS or NIHAO, to further reduce potential biases in the data.

\section*{Broader Impact statement}

The authors are unaware of any immediate ethical or societal implications of this work. This work purely aims to aid in the design of scientific algorithms and datasets. Looking more broadly, the \textbf{GAMMA} dataset and its connection to fundamental physics might help to benchmark new algorithms for image recognition. This may contribute to a broader application and development of interpretable and robust algorithms.   

{
\small
\bibliographystyle{unsrtnat}
\bibliography{references}
}


\end{document}